\pdfoutput=1
\documentclass[aps]{revtex4}
\usepackage{amssymb}
\usepackage{graphicx}
\usepackage{amsmath}
\usepackage{graphicx}
\usepackage{makeidx}
\usepackage{subfigure}
\usepackage{float}

\begin{document}

\title{Symmetric and asymmetric solitons in dual-core couplers with
competing quadratic and cubic nonlinearities}
\author{Lazar Gubeskys and Boris A. Malomed}
\address{Department of Physical Electronics, School of Electrical Engineering,\\
Faculty of Engineering, Tel Aviv University, Tel Aviv 69978, Israel}

\begin{abstract}
We consider the model of a dual-core spatial-domain coupler with $\chi
^{(2)} $ and $\chi ^{(3)}$ nonlinearities acting in two parallel cores. We
construct families of symmetric and asymmetric solitons in the system with
self-defocusing $\chi ^{(3)}$ terms, and test their stability. The
transition from symmetric to asymmetric soliton branches, and back to the
symmetric ones proceeds via a \textit{bifurcation loop}. A pair of stable
asymmetric branches emerge from the symmetric family via a supercritical
bifurcation; eventually, the asymmetric branches merge back into the
symmetric one through a reverse bifurcation. The existence of the loop is
explained by means of an extended version of the cascading approximation for
the $\chi ^{(2)}$ interaction, which takes into regard the XPM par of the $%
\chi ^{(3)}$ interaction. When the inter-core coupling is weak, the
bifurcation loop features a concave shape, with the asymmetric branches
losing their stability at the turning points. In addition to the two-color
solitons, which are built of the fundamental-frequency (FF) and
second-harmonic (SH) components, in the case of the self-focusing $\chi
^{(3)}$ nonlinearity we also consider single-color solitons, which contain
only the SH component but may be subject to the instability against FF
perturbations. Asymmetric single-color solitons are always unstable, whereas
the symmetric ones are stable, provided that they do not coexist with
two-color counterparts. Collisions between tilted solitons are studied too.

OCIS numbers: 190.6135; 190.2620; 230.4320
\end{abstract}

\maketitle

\section{Introduction}

The effect of the second-harmonic (SH) generation by monochromatic light was
first demonstrated in 1961 \cite{Franken}, and has been a topic of great
interest ever since \cite{Torner,New}. The work on this theme includes many
studies dealing with solitons supported by the quadratic ($\chi ^{(2)}$)
nonlinearity \cite{review1,review2}. While the $\chi ^{(2)}$ interactions
between the fundamental-frequency (FF) and SH fields are sufficient for the
creation of solitons, the competition between the $\chi ^{(2)}$ nonlinearity
and its cubic ($\chi ^{(3)}$) counterpart, either self-focusing or
defocusing, may be an essential factor affecting the efficiency of the FF $%
\rightleftarrows $ SH conversion. In addition to the material Kerr
effect, it was predicted \cite{A,F} and experimentally demonstrated
\cite{E,Ady} that $\chi ^{(3)}$ nonlinearity can be
\textit{engineered} by means of the quasi-phase-matching (QPM)
technique, which makes it possible to control the modulational
instability \cite{C} and pulse compression \cite{H} in the medium.
The $\chi ^{(3)}$ nonlinearity may also be induced by semiconductor
dopants implanted into the $\chi ^{(2)}$ material
\cite{semi1}-\cite{semi3}.

Effects of competing $\chi ^{(2)}:\chi ^{(3)}$ nonlinearities on spatial
solitons were analyzed in detail \cite{Komissarova}-\cite{B}. It was
predicted that stable two-color solitons, built of the FF and SH components,
exist in media with the self-defocusing sign of the cubic nonlinearity,
where the solitons do not exist in the absence of the $\chi ^{(2)}$
interactions \cite{Buryak-Kivshar-Trillo-95}. It has also been shown that
the self-focusing $\chi ^{(3)}$ term supports single-color solitary waves
built solely of the SH\ component, but they are prone to destabilization by
the $\chi ^{(2)}$ interactions. In fact, the single-color solitons are
always unstable when they coexist with the two-color ones \cite%
{Bang-Kivshar-Buryak-98}.

The soliton dynamics in symmetric dual-core systems, alias couplers, has
also drawn a great deal of attention, starting from the analysis of solitons
and their bifurcations in the model of twin-core optical fibers with the
intra-core Kerr nonlinearity \cite{fibers1}-\cite{fibers6}, which was
recently followed by the consideration of the soliton stability in the
coupler combining the Kerr terms with the parity-time ($\mathcal{PT}$)
symmetry, represented by equal gain and loss coefficients in the coupled
cores \cite{PT1,PT2}. A similar analysis was reported for spatial solitons
in planar dual-core symmetric waveguides with intrinsic $\chi ^{(2)}$
nonlinearity \cite{Mak-Malomed-Chu-97},\cite{Mak-Malomed-Chu-98}. In all the
cases, the increase of the total energy or power of the soliton (in the
temporal or spatial domain, respectively) leads to the symmetry-breaking
bifurcation (SBB), the difference being that the bifurcation is subcritical
or supercritical in the couplers with the $\chi ^{(3)}$ and $\chi ^{(2)}$
nonlinear terms, respectively. These bifurcations are important examples of
the phenomenology of the spontaneous symmetry breaking of localized modes in
nonlinear media \cite{Springer}.

These findings suggest one to consider dual-core systems with \emph{competing%
} self-focusing and self-defocusing nonlinearities, where the increase of
the total power may, at first, convert symmetric solitons into asymmetric
ones through the SBB, which is followed, at larger powers, by an inverse
symmetry-restoring bifurcation (SRB), if the self-defocusing dominates at
high powers. A natural model of this type is based on the coupler with a
combination of self-focusing cubic and self-defocusing quintic intra-core
nonlinear terms \cite{Albuch-Malomed-07}. The analysis has demonstrated the
existence of the corresponding \textit{bifurcation loops}. A similar effect
was demonstrated in an allied model, based on the nonlinear Schr\"{o}dinger
equation including the combination of the cubic-quintic (CQ) terms and a
double-well potential, which represents a double-channel trapping structure
built into the single waveguide \cite{Zeev}. Dynamical switching between
channels in the latter system was studied too \cite{Radik}. These results
suggest that couplers with competing nonlinearities have a potential for
applications to all-optical data processing.

Saturable nonlinearity is somewhat similar to the combination of the
competing CQ terms. In this context, it is relevant to refer to earlier
works which analyzed the SBB and dynamical switching of solitons in the
coupler with the saturable intra-core nonlinearity \cite{sat1}-\cite{sat4}.
However, that system does not give rise to bifurcation loops, as it does not
feature a switch between the self-focusing and defocusing with the increase
of power.

The above-mentioned $\chi ^{(2)}:\chi ^{(3)}$ combination of interactions is
another natural setting for the realization of the coupler with competing
intra-core nonlinearities. In fact, this setting is more realistic for
experimental realization, as it is not easy to find photonic materials
demonstrating a dominant quintic term without conspicuous losses, while the
self-defocusing $\chi ^{(3)}$ terms may be induced by means of the
above-mentioned techniques. The study of both two- and single-color
solitons, their stability, bifurcations, and interactions in such a system
is the subject of the present work. The model is formulated in Section II.
The main results, in the form of bifurcation diagrams (which feature the
loops) for two-color solitons in the model with the self-defocusing $\chi
^{(3)}$ nonlinearity, are reported in Section III. In addition to systematic
numerical results, we also propose a qualitative explanation of the
existence of the loop, and its collapse which occurs with the increase of
the strength of the linear coupling between the cores, based on the
cascading approximation for the $\chi ^{(2)}$ interaction, to which the $%
\chi ^{(3)}$-induced XPM (cross-phase-modulation) interaction is added. In
Section III, we also consider spontaneous transformation of unstable
solitons into breathers, and collisions between moving (spatially tilted)
stable ones. Results for single-color (SH) solitons in the system with the
self-focusing $\chi ^{(3)}$ terms are produced in Section IV. The paper is
concluded by Section V.

\section{The model}

We consider the co-propagation of the FF and SH fields, $u_{1,2}(x,z)$ and $%
w_{1,2}(x,z)$, in the symmetric coupler (subscripts $1$ and $2$ pertain to
its two cores), with the combined $\chi ^{(2)}:\chi ^{(3)}$ nonlinearity in
its cores. The setting, that does not include losses and spatial walkoff
(which are negligible for relatively short propagation distances available
to the experiment), is described by the system of linearly-coupled
propagation equations in the spatial domain, which can be readily derived as
a combination of those introduced earlier in Refs. \cite%
{Buryak-Kivshar-Trillo-95,G,Bang-Kivshar-Buryak-98,Ady} and \cite%
{Mak-Malomed-Chu-97,Mak-Malomed-Chu-98}:

\begin{eqnarray}
iu_{1z}+u_{1xx}+u_{1}^{\ast }w_{1}+\sigma \left( \frac{1}{4}\left\vert
u_{1}\right\vert ^{2}+2\left\vert w_{1}\right\vert ^{2}\right) u_{1}+Qu_{2}
&=&0,  \notag \\
2iw_{1z}+w_{1xx}-\alpha w_{1}+\frac{1}{2}u_{1}^{2}+\sigma \left( 4\left\vert
w_{1}\right\vert ^{2}+2\left\vert u_{1}\right\vert ^{2}\right) w_{1}+Kw_{2}
&=&0,  \notag \\
&&  \label{eqs} \\
iu_{2z}+u_{2xx}+u_{2}^{\ast }w_{2}+\sigma \left( \frac{1}{4}\left\vert
u_{2}\right\vert ^{2}+2\left\vert w_{2}\right\vert ^{2}\right) u_{2}+Qu_{1}
&=&0,  \notag \\
2iw_{2z}+w_{2xx}-\alpha w_{2}+\frac{1}{2}u_{2}^{2}+\sigma \left( 4\left\vert
w_{2}\right\vert ^{2}+2\left\vert u_{2}\right\vert ^{2}\right) w_{2}+Kw_{1}
&=&0,  \notag
\end{eqnarray}%
Here $\alpha $ represents the phase mismatch between the FF and SH waves, $%
\sigma =+1$ or $-1$ determines, respectively, the self-focusing or
defocusing sign of the $\chi ^{(3)}$ nonlinearity ($|\sigma |=1$ can be
fixed by scaling), while $Q$ and $K$ are inter-core coupling constants for
the FF and SH fields. The total power (norm) of the fields in each core is

\begin{equation}
P_{1,2}=\int_{-\infty }^{+\infty }\left[ \left\vert u_{1,2}(x)\right\vert
^{2}+4\left\vert w_{1,2}(x)\right\vert ^{2}\right] dx,  \label{norm}
\end{equation}%
the conserved power being $P\equiv P_{1}+P_{2}$.

Stationary soliton solutions with propagation constant $\lambda $ are looked
for as

\begin{eqnarray}
\left\{ u_{1}\left( x,z\right) ,w_{1}\left( x,z\right) \right\} &=&\left\{
U_{1}\left( x\right) ,W_{1}\left( x\right) \right\} e^{i\lambda z},  \notag
\\
&&  \label{eqs-stationary} \\
\left\{ u_{2}\left( x,z\right) ,w_{2}\left( x,z\right) \right\} &=&\left\{
U_{2}\left( x\right) ,W_{2}\left( x\right) \right\} e^{2i\lambda z},  \notag
\end{eqnarray}%
where real functions $U_{1,2}(x)$ and $W_{1,2}(x)$ satisfy the following
equations:

\begin{eqnarray}
-\lambda U_{1}+U_{1}^{\prime \prime }+U_{1}^{\ast }W_{1}+\sigma \left( \frac{%
1}{4}\left\vert U_{1}\right\vert ^{2}+2\left\vert W_{1}\right\vert
^{2}\right) U_{1}+QU_{2} &=&0,  \notag \\
-\left( 4\lambda +\alpha \right) W_{1}+W_{1}^{\prime \prime }+\frac{1}{2}%
U_{1}^{2}+\sigma \left( 4\left\vert W_{1}\right\vert ^{2}+2\left\vert
U_{1}\right\vert ^{2}\right) W_{1}+KW_{2} &=&0,  \notag \\
&&  \label{eqs-stationary-substitution} \\
-\lambda U_{2}+U_{2}^{\prime \prime }+U_{2}^{\ast }W_{2}+\sigma \left( \frac{%
1}{4}\left\vert U_{2}\right\vert ^{2}+2\left\vert W_{2}\right\vert
^{2}\right) U_{2}+QU_{1} &=&0,  \notag \\
-\left( 4\lambda +\alpha \right) W_{2}+W_{2}^{\prime \prime }+\frac{1}{2}%
U_{2}^{2}+\sigma \left( 4\left\vert W_{2}\right\vert ^{2}+2\left\vert
U_{2}\right\vert ^{2}\right) W_{2}+KW_{1} &=&0,  \notag
\end{eqnarray}%
with the prime standing for $d/dx$. These equations were solved\ numerically
by means of the Newton-Raphson method based on a finite-difference scheme.
The asymmetry of solitons with unequal amplitudes in the two cores, $\left(
U_{1,2}\right) _{\max }$ and $\left( V_{1,2}\right) _{\max }$, is measured
by
\begin{equation}
\mathrm{\Theta }_{\mathrm{F}}=\frac{\left( U_{1}\right) _{\max }^{2}-\left(
U_{2}\right) _{\max }^{2}}{\left( U_{1}\right) _{\max }^{2}+\left(
U_{2}\right) _{\max }^{2}}.  \label{asymmetry-ratio}
\end{equation}%
This parameter is defined in terms of the FF component only, as its
amplitudes in the two-color solitons are always essentially larger than the
SH amplitudes.

The stability of the soliton solutions was investigated by means of direct
simulations of the perturbed evolution within the framework of Eqs. (\ref%
{eqs}). The simulations were carried out using the fourth-order Runge-Kutta
algorithm.

The bifurcation loops are produced in the next section for $\sigma =-1,$
i.e., the self-defocusing sign of the $\chi ^{(3)}$ nonlinearity. This
choice is necessary, once we aim to consider the competing nonlinear
interactions, $\chi ^{(2)}$ (effectively self-focusing, in the cascading
limit \cite{Torner}) and $\chi ^{(3)}$. Most results will be presented for $%
\alpha =0.5$, as this value of the mismatch adequately represents the
generic situation. However, some essential findings (the critical size of
the coupling, $Q_{\mathrm{cr}}$, at which the bifurcation loop shrinks to
zero) are included too for other values of $\alpha $.

The parameters are different in Section IV dealing with single-color
solitons. In particular, $\sigma =+1$ is fixed in that case, as such
solitons exist only for the self-focusing $\chi ^{(3)}$ nonlinearity.

\section{Bifurcation diagrams and dynamics of two-color solitons}

\subsection{The bifurcation loops}

\begin{figure}[tbp]
\begin{center}
\includegraphics[width=.7\columnwidth]{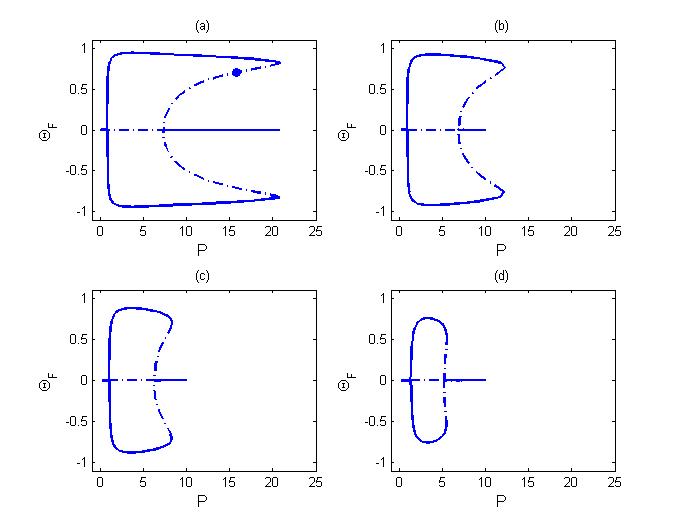}
\end{center}
\caption{(Color online) Concave bifurcation loops for solitons, obtained at
different values of the FF inter-core coupling coefficient $Q$, are
displayed by means of the dependence of asymmetry measure (\protect\ref%
{asymmetry-ratio}) on the total power (norm), $P$: (a) $Q=0.0047$, (b) $%
Q=0.0054$, (c) $Q=0.0065$, (d) $Q=0.0085$. Here and in the next figure, as
well in Fig. \protect\ref{fig6b} below, stable and unstable portions of the
solution branches are shown by solid and dashed-dotted segments,
respectively. The dot on the unstable asymmetric branch indicates a solution
whose spontaneous transformation into a symmetric breather is displayed
below in Fig. \protect\ref{fig3}. This figure and ones following below are
drawn for $\protect\alpha =0.5$.}
\label{fig1}
\end{figure}

\begin{figure}[tbp]
\begin{center}
\includegraphics[width=.7\columnwidth]{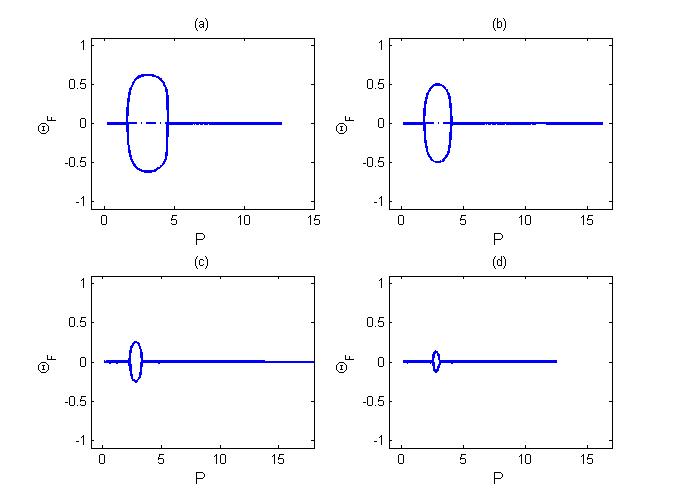}
\end{center}
\caption{(Color online) The continuation of Fig. (\protect\ref{fig1}) --
convex bifurcation loops obtained at larger values of the FF coupling
constant: (a) $Q=0.0098$, (b) $Q=0.0106$, (c) $Q=0.0115$, (d) $Q=0.0117$. In
panels (c) and (d), the short unstable segment of the symmetric solitons
inside the loops is not plotted.}
\label{fig2}
\end{figure}

Because the asymmetry of the solitons produced by the SBB, defined as per
Eq. (\ref{asymmetry-ratio}), is their most important characteristic, the
bifurcation diagrams were generated by varying propagation constant $\lambda
$ and plotting the respective results in the form of $\mathrm{\Theta }_{%
\mathrm{F}}$ [see Eq. (\ref{asymmetry-ratio})] vs. the total power, $P$, at
fixed values of the FF coupling constant, $Q$, and for $K=0$. The latter
assumption is relevant, as the evanescent field, which accounts for the
inter-core coupling, decays faster in the SH component. In fact, results
were also obtained for $K\neq 0,$ showing slight additional shrinkage of the
bifurcation loops displayed in Figs. \ref{fig1} and \ref{fig2}. For
instance, for the same values of $Q=0.0054$ and $\alpha =0.5$, which
correspond to Fig. \ref{fig1}(b), the increase of $K$ from zero to $0.0040$
leads to a decrease of the largest power at which the asymmetric solitons
exist by $\approx 9\%$.

Figures \ref{fig1} and \ref{fig2} demonstrate the bifurcation loops of two
different types: concave and convex, at smaller and larger values of $Q$,
respectively. The increase of $Q$ leads to the shrinkage of the loop, which
vanishes (``collapses"), for $\alpha =0.5$, at $Q_{\mathrm{cr%
}}\approx 0.0117$.

\begin{figure}[tbp]
\begin{center}
\includegraphics[width=.7\columnwidth]{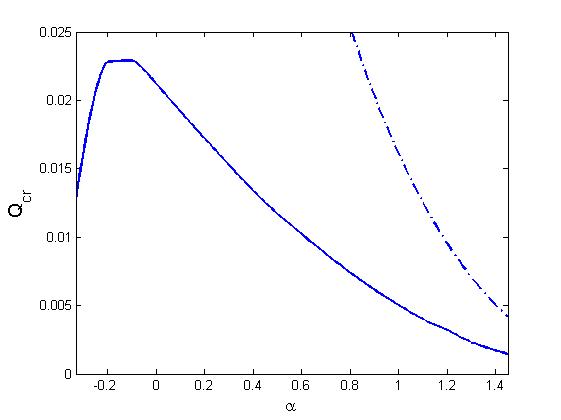}
\end{center}
\caption{(Color online) The solid curve shows the numerically found critical
value of the FF coupling constant, $Q_{\mathrm{cr}}$, at which the
bifurcation loop collapses, versus the phase mismatch, $\protect\alpha $.
The dashed-dotted curve represents the crude analytical approximation for $%
Q_{\mathrm{cr}}$ given by Eq. (\protect\ref{Q}).}
\label{fig:num&anal}
\end{figure}

We have also performed the analysis for other values of the mismatch. The
results are summarized in Fig. \ref{fig:num&anal}, in the form of curve $Q_{%
\mathrm{cr}}(\alpha )$, which is displayed along with qualitative estimate (%
\ref{Q}) derived below.

The concave bifurcation loop connects the symmetry-breaking and restoring
bifurcations of the super- and subcritical types, respectively, while both
bifurcations are supercritical if they are connected by a convex loop. In
agreement with the general principles of the bifurcation theory \cite%
{book,Springer}, segments of the branches of asymmetric solitons in the
concave loop, which connect the point of the supercritical SBB with the
turning points, are stable, while the segments which link the turning points
to the point of the subcritical SRB are unstable. As might be expected too,
the branches of the symmetric solitons, corresponding to $\mathrm{\Theta }_{%
\mathrm{F}}\equiv 0$ in Figs. \ref{fig1} and \ref{fig2}, are unstable inside
the loop, and stable outside.

\subsection{The analytical approximation}

Basic properties of the bifurcation loop can be qualitatively
explained by means of an extended form of the well-known cascading
approximation, which is used to eliminate (``enslave") the SH field
in favor of the FF, when the mismatch is large and the diffraction
term may be neglected in the equations for the SH
\cite{Torner,review1,review2} (a more sophisticated version of the
cascading approximation, which makes use of the
Fourier decomposition of periodic structures, was developed in Refs. \cite%
{A,F,C} for systems based on QPM gratings). This approximation implies that
the amplitude of the SH field is much smaller than its FF counterpart,
hence, applying the approximation to the SH equations in system (\ref%
{eqs-stationary-substitution}), it is necessary to take the XPM term into
account. Thus, the cascading limit yields, in the present case,%
\begin{equation}
W_{n}(x)\approx \frac{\left( U_{n}(x)\right) ^{2}}{2\left[ \alpha
+2\left\vert U_{n}(x)\right\vert ^{2}\right] },~n=1,2,
\end{equation}%
where, as said above, we set $\sigma =-1$. Then, the substitution of this
expression into the FF equations of system (\ref{eqs-stationary-substitution}%
) yields a pair of linearly coupled equations with an effective saturable
nonlinearity:%
\begin{eqnarray}
-\lambda U_{1}+U_{1xx}+\frac{\left\vert U_{1}\right\vert ^{2}}{2\left(
\alpha +2\left\vert U_{1}\right\vert ^{2}\right) }U_{1}-\frac{1}{4}%
\left\vert U_{1}\right\vert ^{2}U_{1}+QU_{2} &=&0,  \notag \\
&&  \label{UU} \\
-\lambda U_{2}+U_{2xx}+\frac{\left\vert U_{2}\right\vert ^{2}}{2\left(
\alpha +2\left\vert U_{2}\right\vert ^{2}\right) }U_{2}-\frac{1}{4}%
\left\vert U_{2}\right\vert ^{2}U_{2}+QU_{1} &=&0.  \notag
\end{eqnarray}

Unlike the system of coupled equations with the cubic nonlinearity \cite%
{fibers1}-\cite{fibers6}, the SBB\ point cannot be found in an exact form in
this model. However, for a crude estimate one may treat it as cubic system
with effective coefficients%
\begin{equation}
\left( \chi _{1,2}^{(3)}\right) _{\mathrm{eff}}=\frac{1}{2}\left[ \frac{1}{%
\alpha +2\left( U_{1,2}\right) _{\max }^{2}}-\frac{1}{2}\right] ,
\label{coeff}
\end{equation}%
taken at the soliton's center. After that, the application of the well-known
exact result for the SBB point in the cubic system \cite{fibers1} to the
system with coefficients (\ref{coeff}) yields and equation which predicts
the peak power of the symmetric solitons at the bifurcation points, $P_{\max
}\equiv U_{\max }^{2}$:%
\begin{equation}
P_{\max }^{2}+\left( \frac{16Q}{3}-1+\frac{\alpha }{2}\right) P_{\max }+%
\frac{8}{3}\alpha Q=0.  \label{quadr}
\end{equation}%
Obviously, too roots of this quadratic equation correspond to the SBB and
SRB points bounding the bifurcation loop. Further, in the framework of the
present approximation, the above-mentioned critical value of the coupling
constant, $Q_{\mathrm{cr}}$, at which the loop collapses, i.e., the SBB\ and
SRB points merge into one, is identified as a value of $Q$ at which the
discriminant of Eq. (\ref{quadr}) vanishes. The latter condition readily
leads to the following result:%
\begin{equation}
Q_{\mathrm{cr}}=\frac{3}{32}\left( \sqrt{2}-\sqrt{\alpha }\right) ^{2}.
\label{Q}
\end{equation}

As seen in Fig. \ref{fig:num&anal}, the analytical result strongly
overestimates the actual values of $Q_{\mathrm{cr}}$ (which is not
surprising, as the analysis is quite coarse), but, nevertheless, the
approximation provides an explanation of the existence of the bifurcation
loop, as well as the trend of $Q_{\mathrm{cr}}$ to decrease with the
increases of $\alpha $ at $\alpha >0$. The decrease of the discrepancy with
the increase of the mismatch, $\alpha $, is a natural feature of the
cascading approximation.

\subsection{Transformations of unstable solitons}

Direct numerical simulations demonstrate that the unstable symmetric
solitons evolve towards breathers oscillating around either of the stable
asymmetric soliton existing at the same total power (not shown here in
detail, as the picture is similar to that observed in previously studied
models). This transition gives rise to very little radiation loss, i.e., the
dynamical symmetry breaking practically keeps the initial total power.

\begin{figure}[tbp]
\begin{center}

\includegraphics[width=\columnwidth]{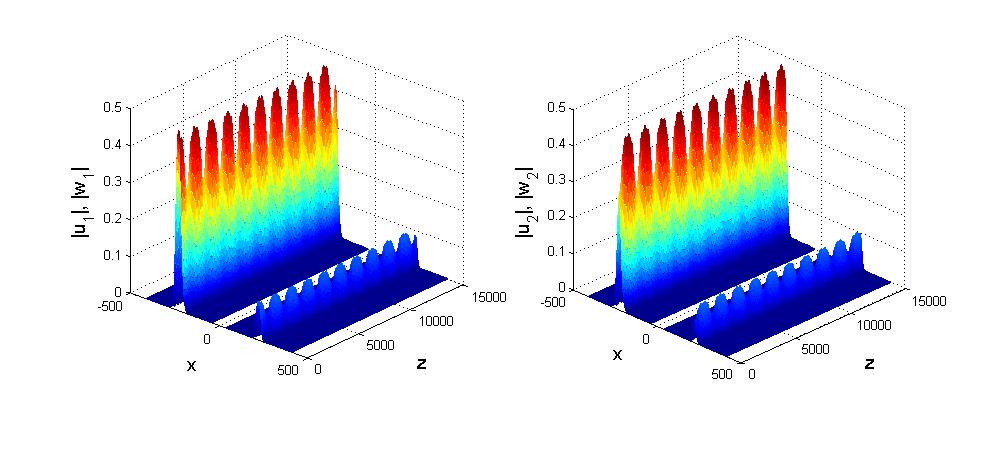}

\end{center}
\caption{(Color online) The evolution, initiated by small perturbations, of
the FF and SH\ components in cores 1 and 2 (left and right panels,
respectively) of the unstable asymmetric soliton marked by the dot in Fig.
\protect\ref{fig1}(a), with $Q=0.0047$, propagation constant $\protect%
\lambda =0.0255$, and total power $P=15.77$. In each panel, the plots of the
FF and SH components are juxtaposed with a shift, for the purpose of clearer
presentation.}
\label{fig3}
\end{figure}

\begin{figure}[tbp]
\begin{center}
\includegraphics[width=.8\columnwidth]{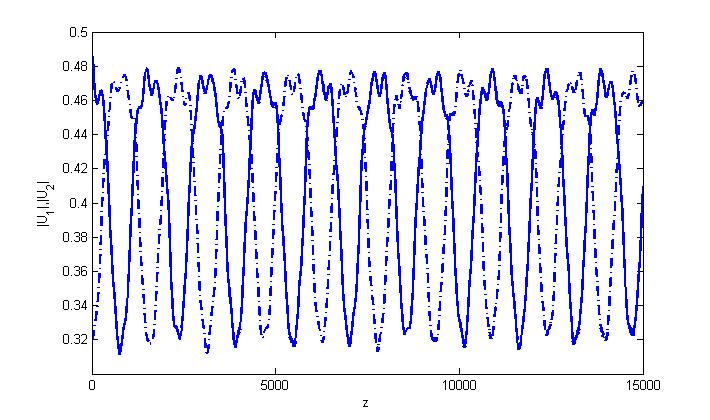}
\end{center}
\caption{(Color online) Periodic out-of-phase oscillations of amplitudes of
the FF fields in the two cores, corresponding to the situation displayed in
Fig. \protect\ref{fig3}.}
\label{fig3_extra}
\end{figure}

More interesting is the spontaneous transformation of asymmetric solitons
belonging to the unstable segments of the concave loop, which is a specific
feature of the present setting, see Fig. \ref{fig1}. An example, displayed
in Fig.~\ref{fig3}, demonstrates that these unstable solitons spontaneously
evolve into breathers, which are symmetric on the average, with the fields
in the two cores oscillating out of phase, as shown in Fig. \ref{fig3_extra}%
. Overall, this effect may be considered as \textit{dynamical
resymmetrization} of the solitons. This transition also conserves the
initial power of the unstable soliton almost exactly, and the emerging
breathers emit virtually no radiation waves.

\subsection{Collisions between two-color solitons}

Equations (\ref{eqs}) are obviously invariant with respect to the spatial
tilt, which is a counterpart of the Galilean boost in the temporal domain:

\begin{eqnarray}
u_{1,2}(x,z) &\rightarrow &u_{1,2}\left( x-cz,z\right) \exp \left( \frac{i}{2%
}cx-\frac{i}{4}c^{2}z\right) ,  \notag \\
w_{1,2}(x,z) &\rightarrow &w_{1,2}\left( x-cz,z\right) \exp \left( icx-\frac{%
i}{2}c^{2}z\right) ,  \label{transformation}
\end{eqnarray}%
where $c$ is an arbitrary real tilt constant. This invariance makes it
natural to consider collisions between spatial solitons tilted in opposite
directions. Direct simulations of the collisions make it possible to draw
the following conclusions.

\begin{figure}[tbp]
\begin{center}
\includegraphics
[width=\columnwidth]{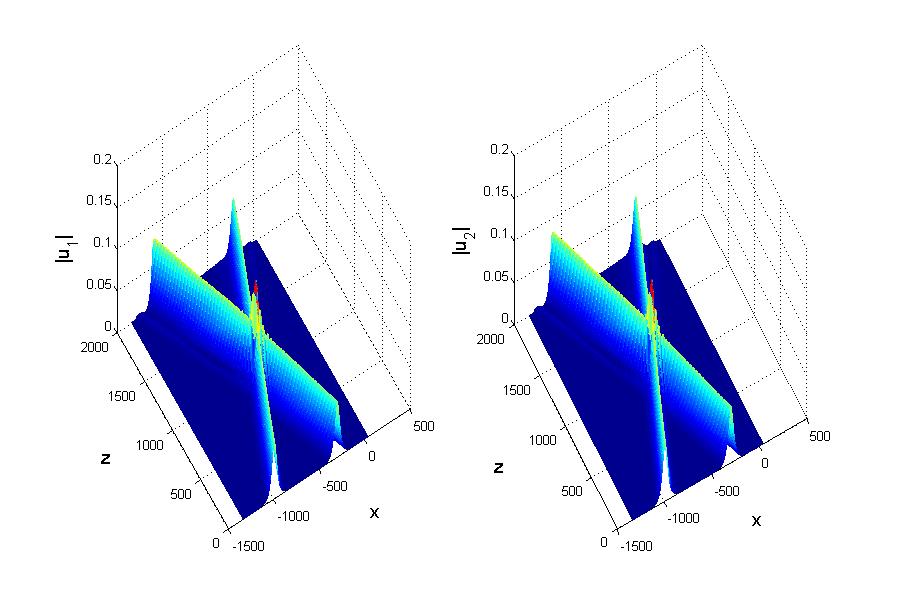}
\end{center}
\caption{(Color online) The elastic collision between low-power solitons,
for $Q=0.003$ and $\protect\lambda =0.0046$ (which corresponds to total
power $P=0.5162$). In this figure and similar ones which display collisions,
the left and right plots represent, respectively, the FF fields in cores 1
and 2. The picture of the SH component is quite similar.}
\label{fig4}
\end{figure}

\begin{figure}[tbp]
\begin{center}
\includegraphics
[width=\columnwidth]{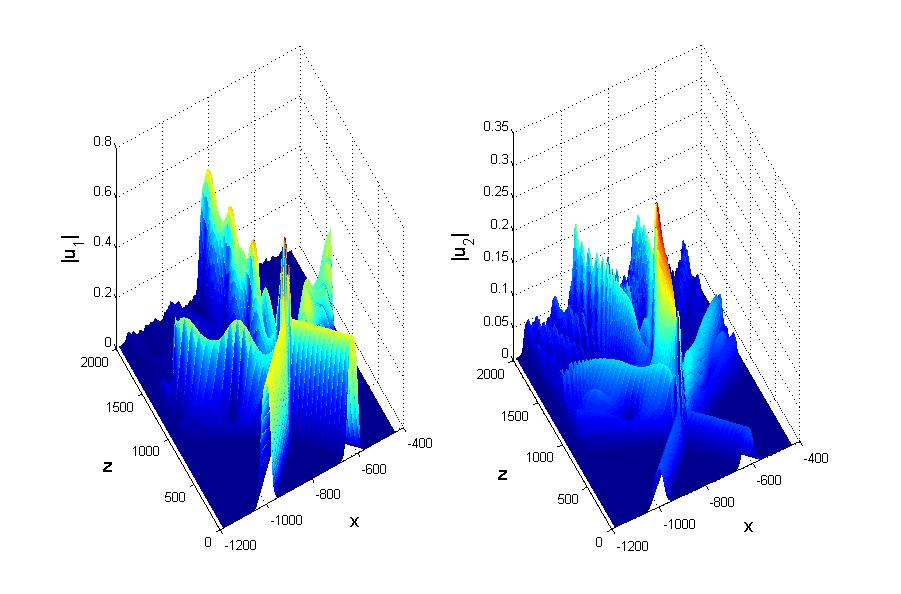}
\end{center}
\caption{(Color online) The inelastic collision between identical high-power
asymmetric solitons, at $Q=0.003$ and $\protect\lambda =0.0248$ (which
corresponds to total power $P=10.43$).}
\label{fig5a}
\end{figure}

\begin{figure}[tbp]
\begin{center}
\includegraphics
[width=\columnwidth]{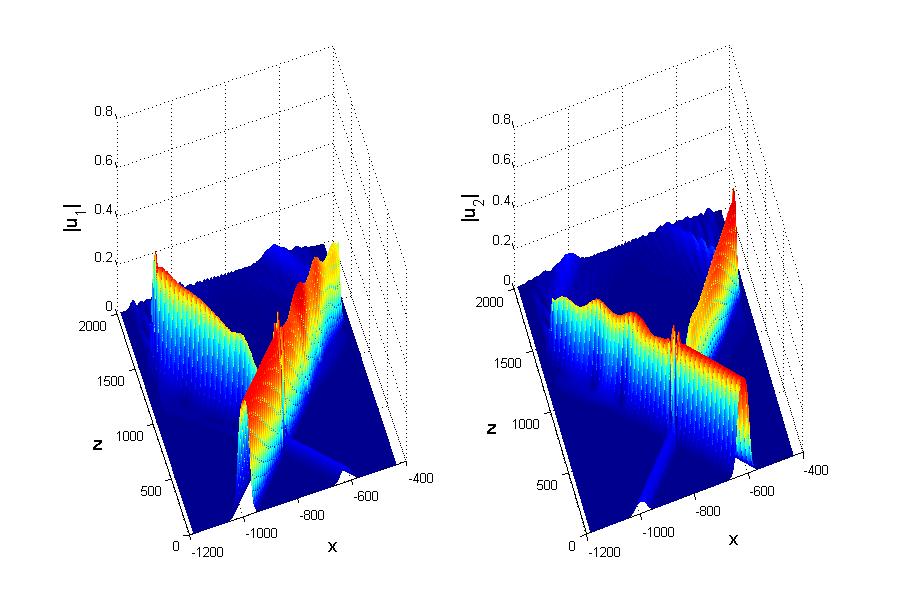}
\end{center}
\caption{(Color online) The inelastic collision between strongly asymmetric
solitons, one being a mirror image of the other (i.e., the larger component
of one soliton collides with the smaller component of the other), for $%
Q=0.003$ and $\protect\lambda =0.0245$, which corresponds to total power $%
P=5.63$.}
\label{fig5b}
\end{figure}

At low powers, the collisions are, as might be expected, nearly elastic, and
do not affect tracks of the\ two beams, see an example in Fig.~\ref{fig4}
for a pair of identical stable symmetric solitons. At higher powers, the
collisions are inelastic. An example, this time for asymmetric solitons, is
shown in Fig.~\ref{fig5a}, where larger components in both solitons belong
to the same core. In this case, the collision leads to splitting of the two
solitons into multiple pulses. In addition, Fig. \ref{fig5b} displays an
example of the inelastic collision between strongly asymmetric solitons
which are specular counterparts to each other, with the larger components
belonging to different cores. In the latter case, the collision tends to
transform the solitons into more symmetric ones.

\begin{figure}[tbp]
\begin{center}
\includegraphics
[width=\columnwidth]{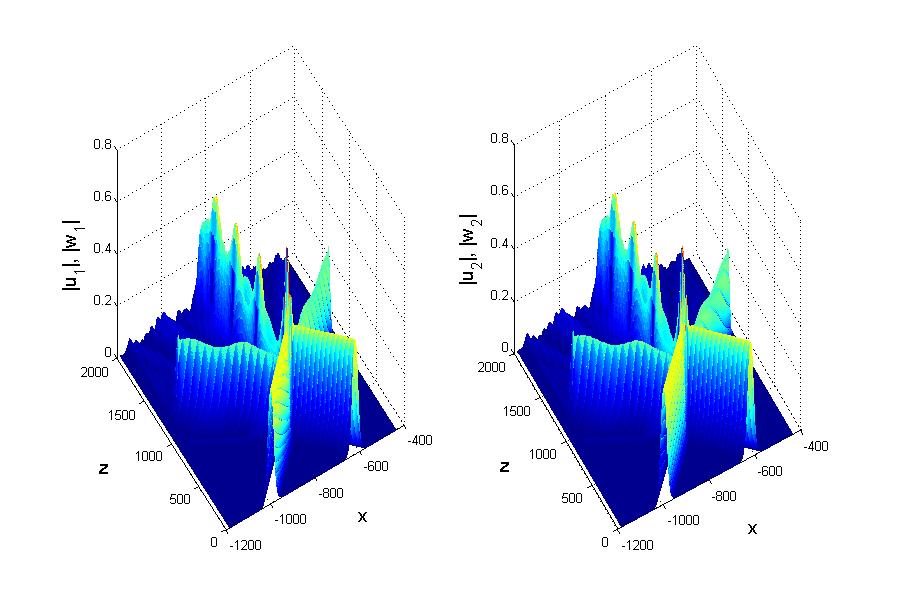}
\end{center}
\caption{(Color online) The inelastic collision between identical symmetric
solitons whose power exceeds the largest power of asymmetric ones, i.e.,
they are located to the right of the bifurcation loop, in terms of Figs.
\protect\ref{fig1} and \protect\ref{fig2}. The parameters are $Q=0.003$, $%
\protect\lambda =0.0267$, and $P=10.99$.}
\label{fig5c}
\end{figure}

Lastly, collisions between high-power symmetric solitons, which are located
to the right of the bifurcation loops, in terms of Figs. \ref{fig1} and \ref%
{fig2}, are strongly inelastic too. An example, displayed in Fig. \ref{fig5c}%
, may be roughly interpreted as transformation of the two colliding solitons
into three, which emerge in strongly excited states, while the spatial
symmetry of the configuration persists after the collision.

\section{Single-color (second-harmonic) solitons}

To allow the existence of single-color SH solitons in the model with the
self-focusing $\chi ^{\left( 3\right) }$ terms, i.e., $\sigma =+1$ in Eqs. (%
\ref{eqs}), it is necessary to include the inter-core coupling of the SH
fields, $K$. Therefore, we here fix $Q=0.2$ and $K=0.5$. In fact, the
particular value of $Q$ is not crucially important in the present context,
where the FF fields are essential for the destabilization of the SH
solitons. This instability is not strongly affected by the linear coupling
between the FF fields in the two cores.

\begin{figure}[tbp]
\begin{center}
\includegraphics
[width=\columnwidth]{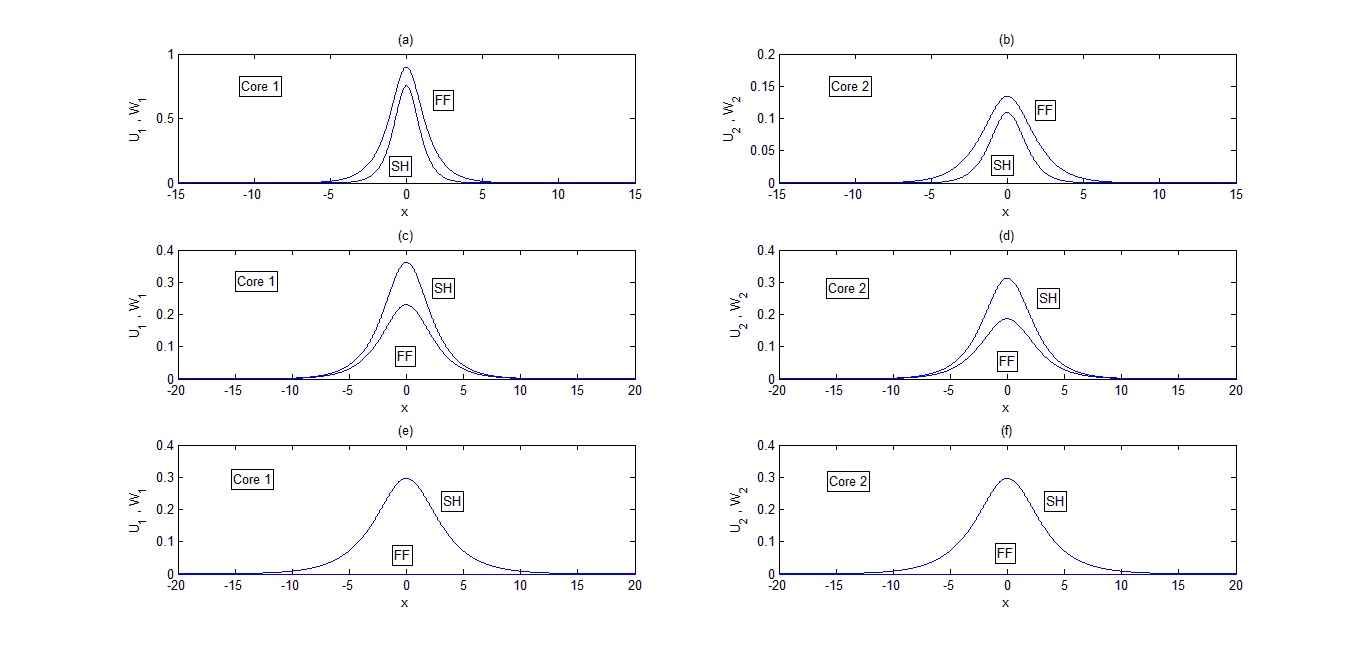}
\end{center}
\caption{(Color online) The transition from two-color solitons in (a,b) and
(c,d), with the dominant FF or SH components, respectively, the to the
single-color (SH)\ soliton in panels (e,f), in the system with the
self-focusing $\protect\chi ^{(3)}$ nonlinearity ($\protect\sigma =+1$),
relatively strong inter-core coupling in the SH field ($Q=0.2$, $K=0.5$),
and mismatch $\protect\alpha =-1.3$. The propagation constants of the
solitons and powers of their components are: $\protect\lambda =0.1$,$~\left(
P_{u}\right) _{1}=1.5688$, $\left( P_{w}\right) _{1}=0.3328$, $\left(
P_{u}\right) _{2}=0.0523$, $\left( P_{w}\right) _{2}=0.0956$ in (a,b); $%
\protect\lambda =0.531$, $\left( P_{u}\right) _{1}=0.2027$, $\left(
P_{w}\right) _{1}=1.8684$, $\left( P_{u}\right) _{2}=0.1413$, $\left(
P_{w}\right) _{2}=1.4836$ in (c,d); and $\protect\lambda =0.494$, $\left(
P_{u}\right) _{1,2}=0$, $\left( P_{w}\right) _{1,2}=1.678$ in (e,f). The
two-color asymmetric soliton is stable in (a,b) and unstable in (c,d). The
single-color symmetric soliton in (e,f) is located at a boundary between
stable and unstable subfamilies.}
\label{fig6}
\end{figure}

Figure~\ref{fig6} demonstrates a set of soliton shapes obtained from Eqs. (%
\ref{eqs-stationary-substitution}), which illustrate the transition from
(asymmetric) two-color solitons [panels (a,b) and (c,d)] to the single-color
symmetric one [panels(e,f)] with the decrease of the total power. The
numerical investigation demonstrates that this transition occurs at $\alpha
>\alpha _{\min }$ $\approx $ $-1.8$. It has been inferred that the
single-color solitons are stable \emph{precisely} at those values of the
propagation constant, $\lambda $ [see Eq. (\ref{eqs-stationary})], at which
neither symmetric nor asymmetric two-color solitons exist, which implies the
absence of the drive for the parametric instability of the SH mode against
excitation of FF perturbations.

\begin{figure}[tbp]
\begin{center}
\includegraphics
[width=.7\columnwidth]{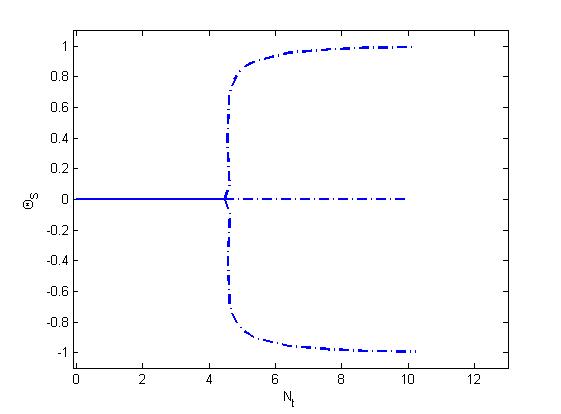}
\end{center}
\caption{(Color online) The symmetry-breaking bifurcation of the
single-color (SH) solitons. Asymmetry $\mathrm{\Theta }_{\mathrm{S}}$ is
defined as per Eq. (\protect\ref{S}). The bifurcation takes place at $%
P\approx 4.47$.}
\label{fig6b}
\end{figure}

At $\alpha <\alpha _{\min }$, single-color solitons appear in an asymmetric
form from their two-color counterparts, also with the decrease of the total
power. All the asymmetric single-color solitons are unstable. With the
further decrease of the total power, they turn symmetric and,
simultaneously, become stable. For example, at $\alpha =-2.32$, the FF
component of the two-color soliton vanishes at $\lambda =0.847$, while the
powers of the SH components in the two cores are $3.742$ and $0.8792$. With
the subsequent decrease of $\lambda $ to $0.7881$, the single-color soliton
becomes symmetric and stable. The latter transformation is illustrated by
the SBB bifurcation for the single-color solitons, which is displayed in
Fig. \ref{fig6b}, with the corresponding asymmetry parameter defined as%
\begin{equation}
\mathrm{\Theta }_{\mathrm{S}}=\frac{\left( W_{1}\right) _{\max }^{2}-\left(
W_{2}\right) _{\max }^{2}}{\left( W_{1}\right) _{\max }^{2}+\left(
W_{2}\right) _{\max }^{2}},  \label{S}
\end{equation}%
cf. definition (\ref{asymmetry-ratio}) for the two-color solitons.

\begin{figure}[tbp]
\begin{center}
\includegraphics
[width=\columnwidth]{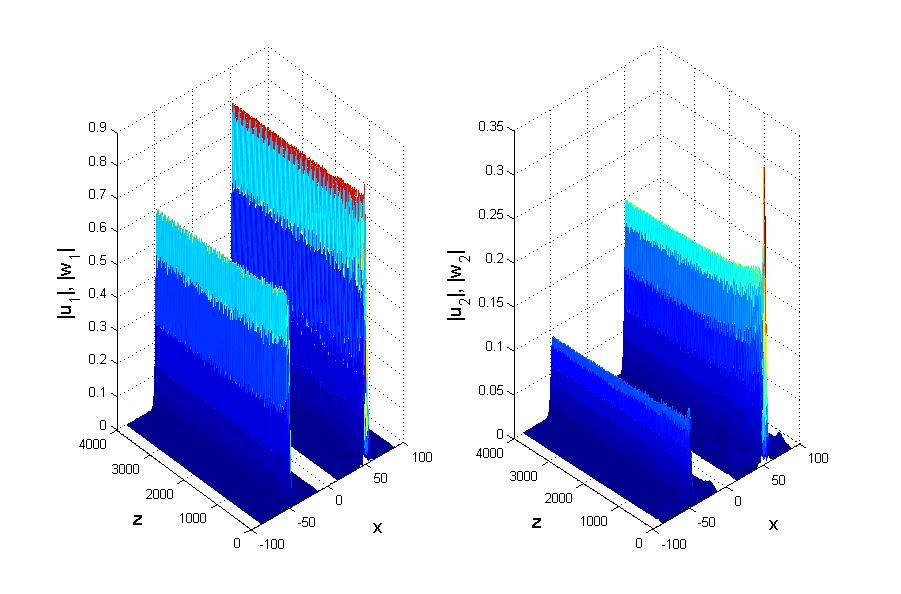}
\end{center}
\caption{(Color online) The evolution of an unstable asymmetric single-color
soliton at $\protect\alpha =-2.32$, with $\protect\lambda =0.8$ and powers
in the two cores $\left( P_{w}\right) _{1}=3.0056$, $\left( P_{w}\right)
_{2}=1.5616$. The left and right panels represent the fields in cores 1 and
core 2, respectively. In each panel, the FF and SH components are shifted to
left and right for clearer presentation.}
\label{fig7}
\end{figure}

An example of the evolution of unstable asymmetric single-color solitons is
displayed in Fig.~\ref{fig7}. The instability results in a very fast
transformation of the soliton into its stable two-color counterpart,
practically without radiative loss of the total power.

\begin{figure}[tbp]
\begin{center}
\includegraphics
[width=.7\columnwidth]{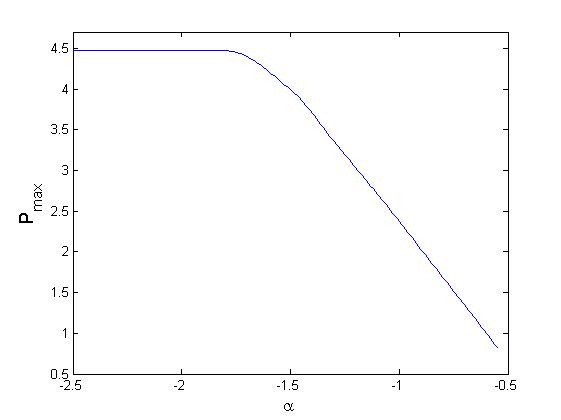}
\end{center}
\caption{(Color online) The SH single-component symmetric solitons are
stable with the power taking values below the maximum value shown here, $%
P<P_{\max }$, which depends on the phase mismatch, $\protect\alpha $.}
\label{fig8}
\end{figure}

Thus, there are two possible causes for the destabilization of the
single-color SH solitons. One is the coexistence with two-color ones, which
triggers the parametric instability against excitation of FF perturbations.
The other instability is caused by the SBB which transforms the symmetric
single-color solitons into asymmetric ones (which themselves are unstable,
as shown in Fig. \ref{fig6b}). A combined stability boundary for the
single-core solitons is displayed in Fig. \ref{fig8}, in which the stability
area is $P<P_{\max }(\alpha )$. The horizontal segment of the boundary
corresponds to the SBB at $P\approx 4.47$, at which the destabilizing
bifurcation happens (see Fig. \ref{fig6b}). Obviously, this critical power
does not depend on $\alpha $, as the phase mismatch amounts to a trivial
shift of the SH propagation constant when the FF field is absent. On the
other hand, the slanted segment with a nearly constant slope accounts for
the parametric instability against the generation of the FF field, when the
single-color soliton coexists with a two-color one. In all the cases, the
evolution of unstable single-color solitons is quite similar to the example
presented in Fig. \ref{fig7}.

\section{Conclusion}

We have introduced the model of the dual-core waveguide with the combined $%
\chi ^{(2)}:\chi ^{(3)}$ nonlinearity acting in the coupled cores. The main
objective was to study the SBB-SRB
(symmetry-breaking-bifurcation---symmetry-restoring-bifurcation) sequence
for two-color solitons in the coupler with the competing $\chi ^{(2)}$ and
defocusing $\chi ^{(3)}$ nonlinearities, which gives rise to the \textit{%
bifurcation loop}. For weaker and stronger inter-core coupling, the loops
feature concave and convex shapes, respectively. They shrink and disappear
with the increase of the coupling constant. The existence and eventual
collapse of the loop can be explained with the help of the cascading
approximation for the $\chi ^{(2)}$ interaction, taking into regard the $%
\chi ^{(3)}$-induced XPM interaction. The stability of the symmetric and
asymmetric solitons follows general principles of the bifurcation theory.
Collisions between tilted solitons show a trend to the transition from
quasi-elastic to inelastic collisions with the increase of the solitons'
power. Further, single-color SH solitons were investigated in the model with
the self-focusing $\chi ^{(3)}$ nonlinearity. Symmetric single-color
solitons are stable, provided that they do not coexist with two-color ones,
up to the SBB point of the single-color solitons.

It may be interesting to extend the analysis for the coupler carrying the
three-wave system, with the Type-II $\chi ^{(2)}$ interaction between two FF
and one SH\ components, competing with the $\chi ^{(3)}$ nonlinearity. A
challenging problem is to extend the analysis for spatiotemporal solitons in
the planar dual-core waveguide with the combined quadratic and cubic
nonlinearities acting in each core.

\end{document}